\documentclass[aps,twocolumn,superscriptaddress]{revtex4-1}

\usepackage{CJK}
\usepackage{graphicx}
\usepackage{epstopdf}
\usepackage{amsmath}
\usepackage{amssymb}
\usepackage{color}
\usepackage{txfonts}
\usepackage{verbatim}
\usepackage{cprotect}
\usepackage{hyperref}

\begin{document}
\begin{CJK}{UTF8}{mj} 

\title{Intellectual interchanges in the history of the massive online open-editing encyclopedia, Wikipedia}


\author{Jinhyuk Yun (윤진혁)}
\affiliation{Department of Physics, Korea Advanced Institute of Science and Technology, Daejeon 34141, Korea}

\author{Sang Hoon Lee (이상훈)}
\email[Corresponding author: ]{lshlj82@kias.re.kr}
\affiliation{School of Physics, Korea Institute for Advanced Study, Seoul 02455, Korea}

\author{Hawoong Jeong (정하웅)}
\email[Corresponding author: ]{hjeong@kaist.edu}
\affiliation{Department of Physics, Korea Advanced Institute of Science and Technology, Daejeon 34141, Korea}
\affiliation{Institute for the BioCentury, Korea Advanced Institute of Science and Technology, Daejeon 34141, Korea}
\affiliation{Asia Pacific Center for Theoretical Physics, Pohang, Gyeongbuk 37673, Korea}


\date{\today}

\begin{abstract}
Wikipedia is a free Internet encyclopedia with an enormous amount of content. This encyclopedia is written by volunteers with various backgrounds in a collective fashion; anyone can access and edit most of the articles. This open-editing nature may give us prejudice that Wikipedia is an unstable and unreliable source; yet many studies suggest that Wikipedia is even more accurate and self-consistent than traditional encyclopedias. Scholars have attempted to understand such extraordinary credibility, but usually used the number of edits as the unit of time, without consideration of real-time. In this work, we probe the formation of such collective intelligence through a systematic analysis using the entire history of $34\,534\,110$ English Wikipedia articles, between 2001 and 2014. From this massive data set, we observe the universality of both timewise and lengthwise editing scales, which suggests that it is essential to consider the real-time dynamics. By considering real time, we find the existence of distinct growth patterns that are unobserved by utilizing the number of edits as the unit of time. To account for these results, we present a mechanistic model that adopts the article editing dynamics based on both editor-editor and editor-article interactions. The model successfully generates the key properties of real Wikipedia articles such as distinct types of articles for the editing patterns characterized by the interrelationship between the numbers of edits and editors, and the article size. In addition, the model indicates that infrequently referred articles tend to grow faster than frequently referred ones, and articles attracting a high motivation to edit counterintuitively reduce the number of participants. We suggest that this decay of participants eventually brings inequality among the editors, which will become more severe with time.
\end{abstract}


\maketitle


\section{Introduction}
Humankind has accrued \textit{a priori} knowledge since the onset of \textit{Homo sapiens}. From ancient cave paintings to modern research papers, the species' desire toward sedimentation has been displayed as a documentary. An encyclopedia, a set of documents that contains a vast collection of information from the entire field of human knowledge, has played a pivotal role in disseminating these legacies~\cite{Pliny77,Lindberg2007}. Conventionally, a group of experts devote their expertise to these encyclopedias~\cite{Britannica, Americana}. Taking advantage of technological developments, media that publish encyclopedias keep abreast of the times: handwriting, letterpress printing, and optical disks. The emergence of information technology has opened a new era of publishing traditional encyclopedias on the World Wide Web~\cite{BritannicaWeb}, which offers a variety of references and up-to-date information. Although these new media can reduce the publication price, encyclopedia editing is still costly. 
 
Besides the improvement of traditional encyclopedias, new media enable fresh challengers to participate in the competition. Wikipedia~\cite{Wikipedia}, a representative player among the challengers, has proposed an entirely new manner: editing by volunteers with various backgrounds in a collective fashion. This new paradigm of sharing knowledge is one of the most famous examples of ``collective intelligence.'' However, due to the nature of open-edit policy, Wikipedia does not guarantee that the contents are valid~\cite{WikipediaDisclamer}, thus it is regarded ambiguous and even inaccurate to utilize in scientific context~\cite{Kittur2008, Adler2008}. Despite such a long-standing bias against the credibility of Wikipedia, many studies suggest that Wikipedia is more reliable than our prejudice; Wikipedia itself tends to refer reliable scientific sources~\cite{Nielsen2007, Haigh2011}. Only 13\% of Wikipedia articles contain perceptible academic errors~\cite{Chesney2006} and the quantity of factual errors, omissions, or ambiguous statements in scientific context of Wikipedia is comparable to traditional encyclopedias~\cite{Giles2005}.

Gradually, prejudice against the quality of Wikipedia's articles has been eroded and the number of citations to Wikipedia in peer-reviewed scientific articles has increased over time~\cite{Bould2014}. A bizarre gap between such prejudice and the actual situation appeals to the scholars, who have analyzed Wikipedia's external characters and internal dynamics. For example, researchers have investigated editors of Wikipedia and their editing patterns~\cite{Amichai2008, Iba2010, Gandica2015}, and the occurrence and resolving of conflicts in Wikipedia~\cite{Yasseri2012, Iniguez2014}. Despite the significant contributions of such endeavors, the previous studies mainly focus on the raw number of edits, and often neglect real time and the different editing patterns for articles with different sizes and ages. 
In this paper, we examine an exhaustive set of English Wikipedia articles to understand how the article size and age displays external appearance in this open-editing encyclopedia. In particular, a simple time-rescaling method reveals articles belonging to various types, when we take account of the interrelation between observable parameters: the number of edits, the number of editors, and the article size. 

Our analysis consists of both data analysis and modeling based on it. First, we use the entire edit history in Wikipedia to inspect Wikipedia's growth, mainly focusing on the number of edits, the number of editors, and the article size. In this process, we demonstrate that the consideration of real time is essential to understand the underlying dynamics behind the present Wikipedia. Second, to consider the formation of current Wikipedia in more detail, we develop an agent-based model that imitates the interplay between an article and the editors in a society. Our model shows inherent differences of articles belonging to different types of growth patterns. The results are consistent with real data, which suggests that a society's attitudes on Wikipedia articles determine the growth pattern. We believe that this approach provides valuable insights for the formation of collective knowledge. We focus on the long-term formation of collective knowledge, which has significant effects on the progress of humankind over a variety of temporal scales. We hope that our work provides insights to solve some of the fundamental questions: why people collaborate, how the collective memory is formed, and how knowledge is spread and descended.

The rest of the paper is organized as follows. In Sec.~\ref{sec:data_set}, we introduce the Wikipedia data that we use in our investigation. In Sec.~\ref{sec:data_analysis}, we propose a time-rescaling method and show that the articles in Wikipedia can be classified into multiple types based on their growth pattern. We present our model and results in Sec.~\ref{sec:model}, including verification of our model with real-data. Finally, we present our conclusions and discussion in Sec.~\ref{sec:conclusion}.

\begin{figure}
\includegraphics[width=0.5\textwidth]{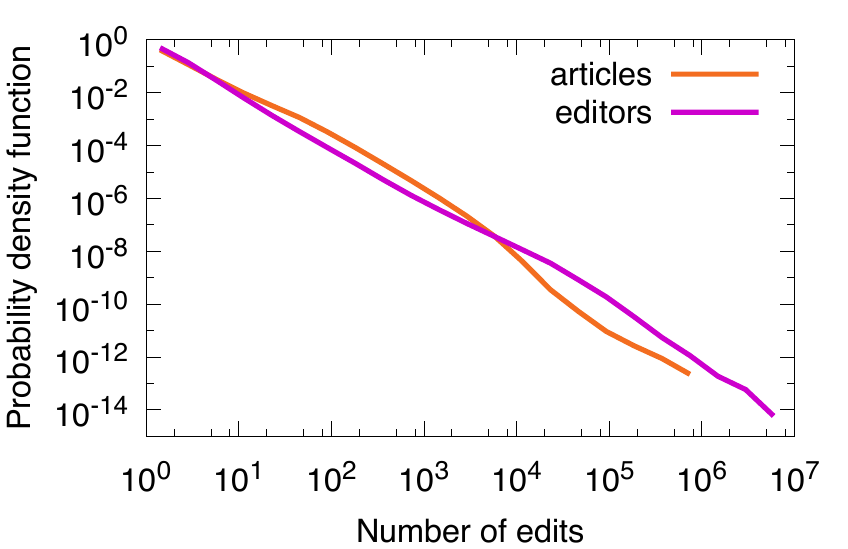}
\caption{Probability distribution of the number of edits for articles, and that for editors (the number of edits in which editors are involved). Both distributions are fat tailed, which implies that even editors and articles with small numbers of edits should not be neglected.}
\label{NumeditsPDF}
\end{figure}

\begin{figure}
\includegraphics[width=0.5\textwidth]{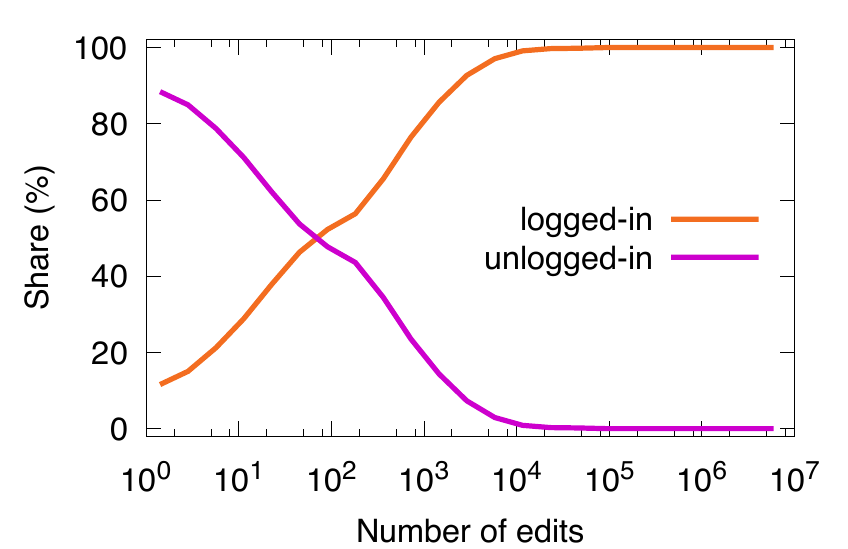}
\caption{Share of logged-in editors and unlogged-in editors in Wikipedia counted based on their number of edits. We observe that most of the heavy editors (who edit repetitively and massively) are logged-in editors, although only 16.1\% ($6\,529\,461$) of editors are logged-in and 83.9\% ($34\,006\,990$) remain anonymous.
}
\label{NumeditsLogged}
\end{figure}

\section{Data Set}
\label{sec:data_set}
For the analysis, we use the December 2014 dump of English Wikipedia. This dump contains the complete copy of Wikipedia articles from the very beginning up to December 8, 2014, including the raw text source and metadata source in the Extensible Markup Language (XML) format~\cite{WikimediaDownloads}. In this data set, there are a total of $34\,534\,110$ articles across all categories with the full history of edits. Each article documents either the Wikipedia account identification (ID) or Internet protocol (IP) address of the editor for each edit, the article size and timestamp for each edit, etc. A single character in English takes 1 byte, so the article size is the direct measure of article length~\cite{Utf8}. There are $587\,606\,564$ editing events (``edits'' from now on) for all Wikipedia articles in total, where individual articles' edit numbers range from $1$ to $1\,044\,048$. 

Previous studies tend to sample data sets for various reasons, and thus articles with small numbers of edits are necessarily filtered out~\cite{Kittur2008, Adler2008, Yasseri2012, Chesney2006}. However, Fig.~\ref{NumeditsPDF} suggests a fat-tailed distribution for the number of edits, so the majority of articles are not edited as many times as the articles in the tail part of the distribution and those articles should not be neglected. Therefore, we consider all entries and use the entire set for analysis. Additionally, we use the ID and IP address, for logged-in editors and unlogged-in editors, respectively, to identify distinct editors. In total, $40\,536\,451$ editors have participated in the establishment of the current Wikipedia. Among them, 83.9\% ($34\,006\,990$) of editors are unlogged-in and only 16.1\% ($6\,529\,461$) of editors are logged-in. Interestingly, the absolute share of logged-in editors is rather smaller than that of unlogged-in editors; most of the heavy editors tend to be logged-in (Fig.~\ref{NumeditsLogged}). Specifically, logged-in editors have modified the articles 455\,397\,682 times (77.5\%) in total, meanwhile unlogged-in editors have modified the articles only 132\,475\,517 times (22.5\%). Considering the fact that the number of unlogged-in editors exceeds that of logged-in editors, the average influence of a single unlogged-in editor is much smaller than that of logged-in editors (on average, 69.8 times per logged-in editor and 3.9 times per unlogged-in editor).

There are possible biases for IP addresses when an IP address is shared, e.g., editors who use a library, public WiFi, virtual private network (VPN), etc., or move the locations. In those cases, there will be under- or overestimation of the number of distinct editors. Additionally, several home Internet connection methods allocate IP addresses dynamically, e.g., digital subscriber line (DSL) and dial-up. For those cases, there might be overestimation and misidentification of distinct editors. However, it is reported that cable and fiber to the home (FTTH) dominate the U.S. market share~\cite{BroadbandShare}, which provide quasistatic IP addresses~\cite{YXie2007}. Considering both the market shares and modest impact of single unlogged-in editors on the current Wikipedia, we believe that our analysis is robust. In fact, we actually check that even when we exclude unlogged-in editors, our results reported in Sec.~\ref{sec:data_analysis} are not affected at all indeed. In addition, a small number of edits does not specify the editor, yet we use other information even for such cases based on the article size and timestamp.

\begin{figure}
\includegraphics[width=0.5\textwidth]{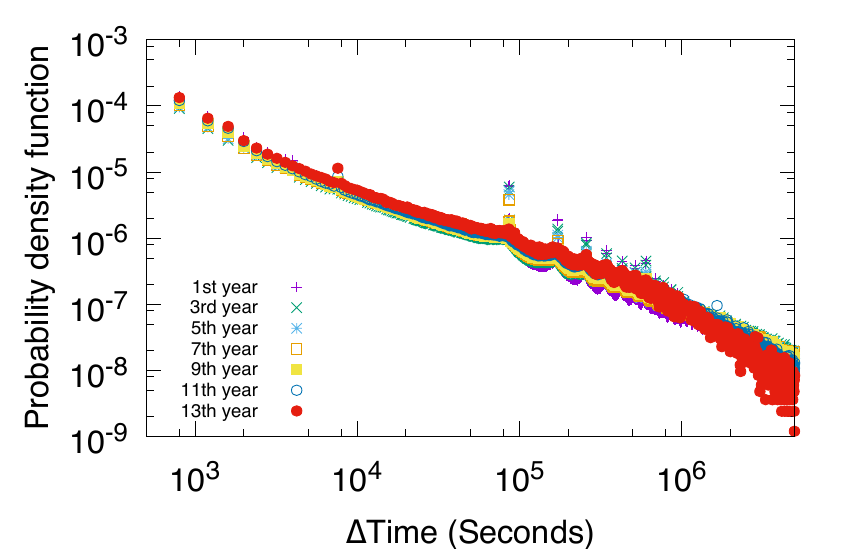}
\caption{Distribution of time between two consecutive edits in Wikipedia. Each symbol corresponds to the range of ages to which articles belong. Specifically, ``$n$-th year'' corresponds to the edit event occurring between the first and the last day of the $n$-th year since the onset of the article. The time differences follow fat-tailed distribution, which is a sign of the burstiness, with a daily periodic pattern ($1$ day = $86\,400$ s).}
\label{DeltatimeperAge}
\end{figure}

\begin{figure}
\includegraphics[width=0.5\textwidth]{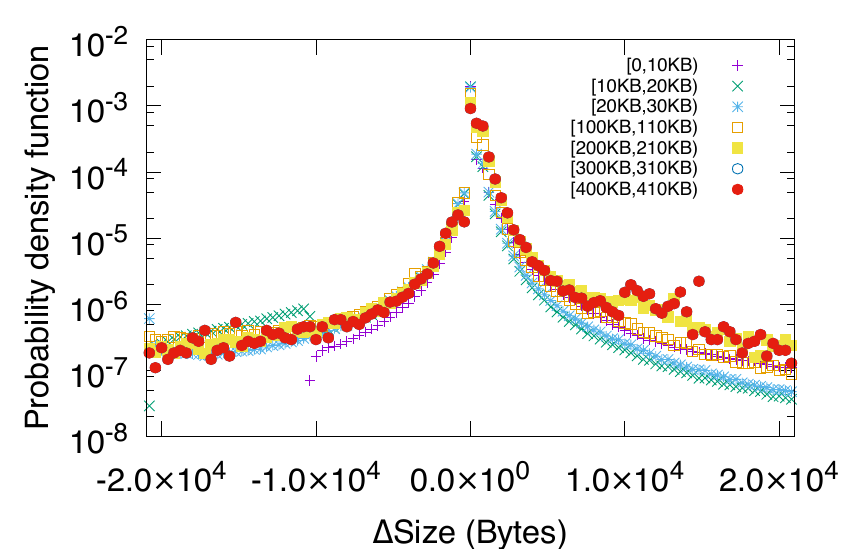}
\caption{Size difference between two consecutive edits in Wikipedia, in the unit of kilobyte (KB). Each symbol corresponds to the range of articles' size when the latter of two consecutive edits occurred.
}
\label{DeltasizeperSize}
\end{figure}

\section{Data Analysis}
\label{sec:data_analysis}

Previous studies on the Wikipedia data set did not use the information about article size changes after the edits~\cite{Yasseri2012, Iniguez2014} or real timestamps of the edits~\cite{Amichai2008, Iba2010, Gandica2015}. We combine such information together with conventional measures, such as the number of edits and the number of editors, to display the nature of Wikipedia. Our first analysis of time and size differences between two consecutive edits reveals regularity, regardless of an article' age and size.

\subsection{Edit scale of Wikipedia}
\label{sec:edit_scale_of_Wikipedia}
The time between the consecutive edits follows a fat-tailed distribution with characteristic periodicity from the human circadian rhythm (Fig.~\ref{DeltatimeperAge}), which suggests that the editing timescale of Wikipedia is intermittent or ``bursty,'' meaning that brief but intense activities are followed by much smaller activities for a long time~\cite{Goh2008, Barabasi2005}. These intense activities in Wikipedia are reported as ``Wikipedia Edit War,'' which refers to significantly rapid consecutive editing by various editors with conflicting opinions~\cite{Kittur2007}. Our observation indicates that the ``edit number'' (or the number of edits), which many studies use as the proxy of the real time~\cite{Iniguez2014, Gandica2015}, is not an unbiased proxy of the time. Counterposed to the assumption that English Wikipedia has already become global media, we observe strong periodicity for the time between the consecutive edits in Fig.~\ref{DeltatimeperAge}. The peaks are located at every $86\,400$ s or a single day, which implies that native English speakers (mostly people in the United States because of the relative population, we presume) still dominate English Wikipedia even though there is no barrier to global access. Such a circadian pattern in the frequency of editing events is mainly driven by editors with specific cultural backgrounds for the data until the beginning of 2010s, as reported in Ref.~\cite{Yasseri2012a}. Our observation indeed shows that the circadian rhythm also affects the interediting time in a collective fashion, and this domination still remains in the current Wikipedia.

Besides the time scale, we observe that an article's growth is mainly addition and subtraction with a characteristic size scale, which are rather independent of the current size (Fig.~\ref{DeltasizeperSize}). This observation is counterposed to the recent report that the growth of collaborated open-source software and mammalian body masses are proportional to their size~\cite{Gherardi2013}, and implies that the influence of a single edit becomes smaller as article size is increased. Most previous research~\cite{Gandica2015, Yasseri2012, Iniguez2014} does not take into account the degree of the influence for a single edit, and thus considers all of the edits as affecting the article of Wikipedia equally. However, our observations propose the necessity of combining the time and size difference between the edits with the conventional measures. 

\begin{figure*}
\includegraphics[width=1.0\textwidth]{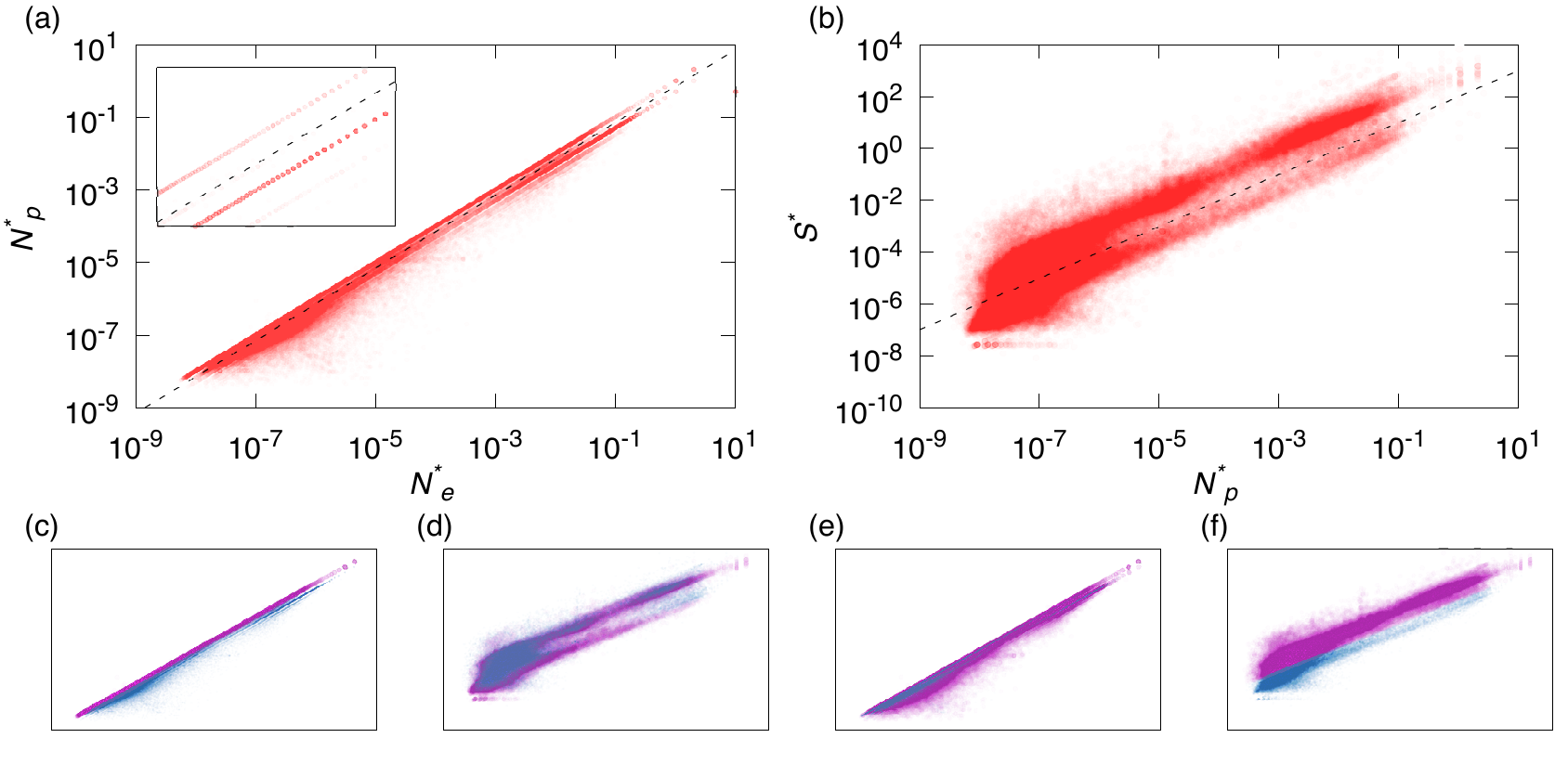}
\caption{Interplay between time-rescaled measures for Wikipedia in December 2014. Each point corresponds to an individual Wikipedia article. For panels (a)--(f), we randomly sampled one article per every 100 articles to display clearly, yet results do not change by sampling. Note that bimodality is observed only when we check these time-rescaled measures. (a) The number of edits ${N}_{e}^{*}$ versus the number of editors ${N}_{p}^{*}$. (b) The number of editors ${N}_{p}^{*}$ and the article size ${S}^{*}$. The black dashed guidelines in panels (a) and (b) have the slope $=1$ ($N_p^{*} = 0.7 N_e^{*}$ and $S^{*} = 100 N_p^{*}$), implying the linearity. The inset in panel (a) is the magnified view on $x \simeq 10^{-1}$ and $y \simeq 10^{-1}$ for visibility. The bimodal structure in (a) is highlighted in (c) by manually dividing the upper (purple) and lower (blue) modes via the black dashed guideline in (a), but this division of modes does not divide the modes of (b), as shown in (d) where we show the same plot as (b) but with the individual points' color of (c). Likewise, the bimodal structure in (b) is highlighted in (f) by manually dividing the upper (purple) and lower (blue) via the black dashed guideline in (b), but this division of modes does not divide the modes of (a), as shown in (e) where we show the same plot as (a) but with the individual points' color of (b). The axes of (c) and (e) are identical to (a) and those of (d) and (f) to (b), so we omit the axes in (c)--(f) for better visualization. Those modality structures qualitatively remain the same even if we exclude the number of edits and the number of editors belonging to unlogged-in editors, from which we conclude that the possible ambiguity in the usage of IP addresses for editor identification can be neglected.
}
\label{Bimodality}
\end{figure*}

\subsection{Time-rescaled measures for Wikipedia}
\label{sec:time_rescaled}

In Sec.~\ref{sec:edit_scale_of_Wikipedia}, we have shown that the time between two consecutive edits is quite heterogeneous (Fig.~\ref{DeltatimeperAge}). This global effect of various timescales itself makes it unfair to directly compare the characteristic parameters of articles: the number of editors, edits, and the article size for different articles. To compensate for such an effect, we employ rescaled measures for article $i$ as ${N}_{e}^{*}(i) = {N}_{e}(i)/{T(i)}$, ${N}_{p}^{*}(i) = {N}_{p}(i)/{T(i)}$, and 
${S}^{*}(i) = {S}(i)/{T(i)}$, where $T(i)$, the age of article $i$, is measured as the time between the moment of onset and that of the latest edit of article $i$. ${N}_{e}(i)$, ${N}_{p}(i)$, and $S(i)$ are the number of edits, the number of editors, and the article size for article $i$, respectively. The rescaled measures are free from the temporal effects, making it possible to recruit myriad articles into the same ground for analysis in the sense of growth per unit time. For the number of edits, the number of editors, and the article size from the data, we hereafter use their rescaled values unless stated otherwise.

A natural step forward is to search for any possible interplay between ${N}_{e}^{*}$, ${N}_{p}^{*}$, and $S^{*}$ 
 in the formation of current Wikipedia. One can suppose that the number of edits has varied gradually as a function of the number of editors, because both measures reflect the degree of interest in the article. Unexpectedly, we discover that the articles show a peculiar bimodality in their number of editors across the entire value of the number of edits [see Fig.~\ref{Bimodality}(a)]. The bimodality is characterized by the linear relation ${N}_{p}^{*} = k_1 {N}_{e}^{*}$ with two distinct proportionality constants, $k_1 = k_1^{\mathrm{upper}} \simeq 0.9$ and $k_1 = k_1^{\mathrm{lower}} \simeq 0.5$, respectively. In other words, there are two groups of articles, determined by the proportion between the number of editors involved in the articles and the editors' average activity; one group is dominated by a relatively small number of enthusiasts who edit articles frequently, and the other is composed of a relatively large number of editors who seldom edit articles. 
Besides the cases of edits and editors, Wikipedia shows a similar division of article size for given numbers of editors [see Fig.~\ref{Bimodality}(b)]. There are two types of articles determined by the average article size produced by an editor per unit time. This relation is also described by the linear dependency $S^{*} = k_2 {N}_{p}^{*}$, where $k_2 = k_2^{\mathrm{upper}} \simeq 900$ for the upper mode and $k_2 = k_2^{\mathrm{lower}} \simeq 25$ for the lower mode. In other words, editors for some articles have generated about $900$ bytes on average, meanwhile the editors of the rest of the articles have generated only about $25$ bytes on average.

\begin{figure*}
\includegraphics[width=1.0\textwidth]{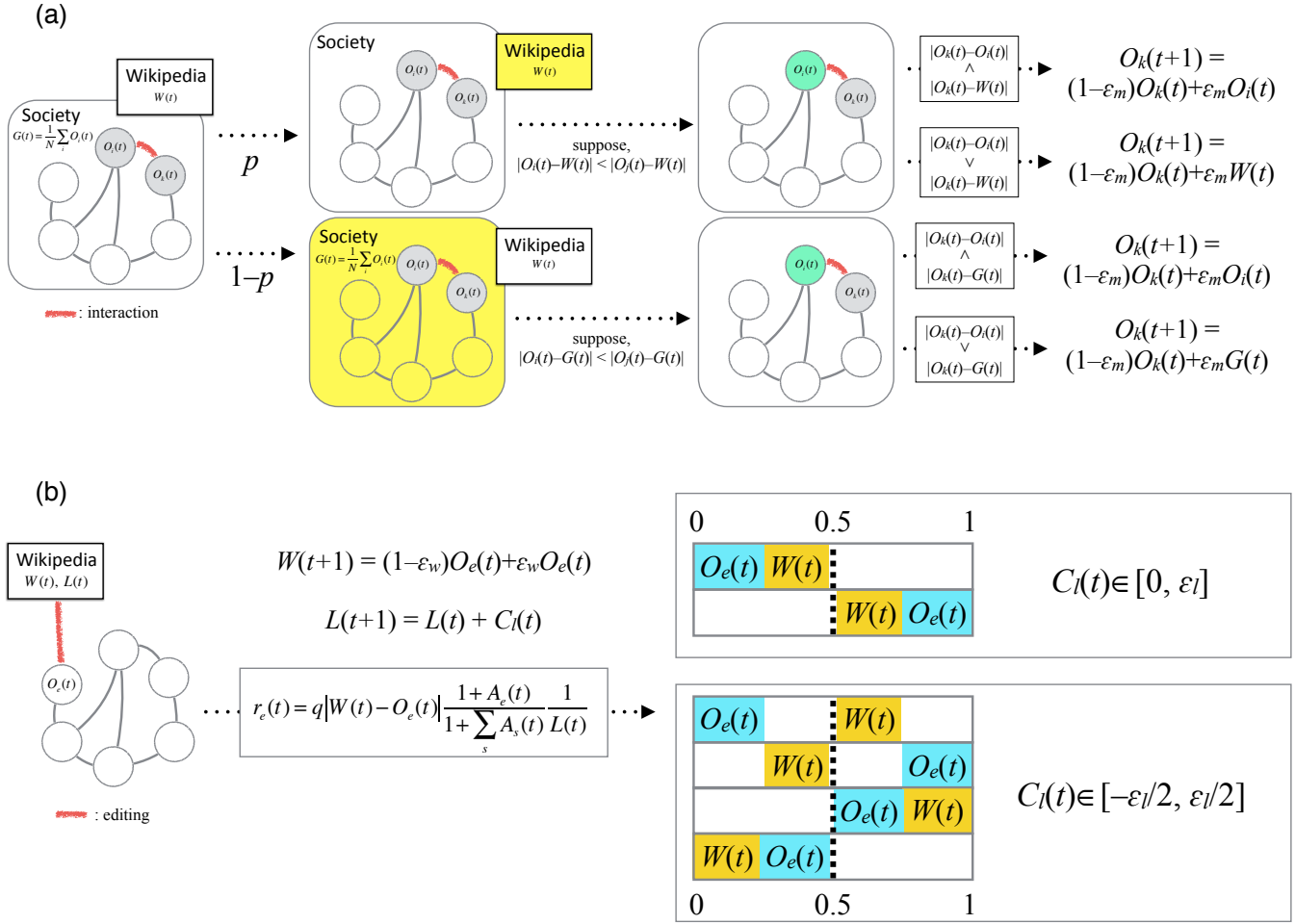}
\caption{Schematic diagrams describing how our model works. (a) Flow diagrams of the interaction between agents $i$ and $k$ described in Sec.~\ref{sec:agent_agent_dynamics}. (b) Flow diagrams of the interaction between agent $e$ and the article described in Sec.~\ref{sec:agent_wiki_dynamics}.
}
\label{Modeldescription_fig}
\end{figure*}
 
Our finding of bimodality in the two relations (${N}_{e}^{*}$ versus ${N}_{p}^{*}$ and ${N}_{p}^{*}$ versus ${S}^{*}$) triggers an interesting question: does each of the modes in one relation correspond to each mode in the other relation [Figs.~\ref{Bimodality}(a) and \ref{Bimodality}(b)]? It seems natural to speculate that such modes have the counterparts in the other relation, or at least one is subordinative of the other. Contrary to this speculation, our observation suggests that there is no visible relationship between the two different types of bimodality [see Figs.~\ref{Bimodality}(c)--\ref{Bimodality}(f)]. The points in the figures are colored according to the modes to which the corresponding articles belong in the criteria based on ${N}_{p}^{*} / {N}_{e}^{*}$ or ${S}^{*} / {N}_{p}^{*}$. We simply tear off the upper and lower modes by drawing a line [the dashed lines in Figs.~\ref{Bimodality}(a) and \ref{Bimodality}(b)] between the two modes and assign purple and blue colors for the points in the upper and the lower modes, respectively. Those purple and blue points are totally mixed when the criterion is based on the other parameter relation. Taken together, we conclude that there are at least four different groups of articles, which can be categorized by its growth per unit time. The possible mechanism behind the division is suggested based on our modeling study in Sec.~\ref{sec:model}.

\section{Mechanistic model of Wikipedia}
\label{sec:model}
To understand the underlying dynamics of the observed patterns, we develop a mechanistic model of editing dynamics by identifying two key factors that drive the evolution of Wikipedia articles. We assume that there are two fundamental and inherent properties of an article reflecting the society's viewpoint on the article's topic: the preferences for referring Wikipedia and the desires to edit (namely, editability). In this section, we show that two such key drivers have elicited the Wikipedia into its current state as shown in Fig.~\ref{Bimodality}. Interestingly, each of those has a decisive effect on the distinct modality structure of ${N}_{p}^{*} / {N}_{e}^{*}$ and ${S}^{*} / {N}_{p}^{*}$, respectively, and they have almost no impact on each other's modality.

The preferences for referring Wikipedia stems from its relative credibility compared to other conventional media. In other words, people tend to refer Wikipedia more than other conventional media or opinion from others for certain topics. Because of the nature of open-edit policy, there are long-lasting arguments of credibility, especially for the scientific contexts~\cite{Kittur2008, Adler2008}. As a result, people avoid referring Wikipedia to reinforce their contention for scientific topics when they debate. Nevertheless, several topics are almost free from the trust issue and Wikipedia can be considered as a trustworthy source of knowledge. The subcultures such as animations, movies, and computer games are good examples, because the editors are not only a fan of the topic but also the creators of such cultures~\cite{Taneska2009}. In those cases, therefore, members of a society do not hesitate to utilize Wikipedia as their grounds for the arguments. 

In addition, there are different levels of psychological barriers and desires in editing, depending on the topic. People tend to edit the article about which they have enough knowledge~\cite{Bryant2005}. Thus, the average ``editability'' of articles, for members of a society, is diverse by its nature from the casual ones which are easily editable to the formal ones. This editability also depends on collective motives, which describe the significance of the topic as the common goal of social movements~\cite{Finkel1998, Lakhani2005}. Therefore, the intrinsic rate of edit should be taken into account. Besides these two key factors, editors are also engaged in articles when they have already given more effort to the articles by editing them~\cite{George2004}, representing the feeling of attachment. Additionally, it is hard to edit an article that already has a massive amount of information~\cite{Bryant2005}, so the motivation to edit will be reduced as the article size is increased. We describe how we implement the sociopsychological effects into our mathematical model in detail. 

\subsection{Model Description}
By incorporating the aforementioned factors, we create a mechanistic model of the article growth. The model comprises $N$ agents where the individual agents represent members of a society and all of the agents are connected to a single Wikipedia article. Note that we take a single Wikipedia article in our model, as we assume that different degrees of editability and credibility yield different types of articles in real Wikipedia. To account for the modality shown in the interplays between three measures, ${S}^{*}$, ${N}_{p}^{*}$, and ${N}_{e}^{*}$ in Fig.~\ref{Bimodality}, we introduce corresponding model parameters. First, the article has its own length $L(t)$ corresponding to ${S}$ in our data analysis. At the beginning, the length is assigned as $L(t=0) = 1$, where $1$ is the minimum length to which agent can reduce the article, so $L(t) \ge 1$ always. The number of edits at time $t$, denoted as $N_e(t)$, is also defined as the total number of article updates until $t$, under the update rules described in Sec.~\ref{sec:agent_wiki_dynamics}. Additionally, $N_p(t)$ corresponds to the number of distinct agents who edited the article at least once. 

Besides the quantities explicitly measured in data analysis, we also adopt internal parameters for the agents and the article. The agents are connected to each other with the Erd\H{o}s-R{\'e}ny random network~\cite{Erdos1960, Gilbert1959}. Such connections between agents stand for various relationships in society: friends, co-workers, even enemies. Every agent $i$ has its own opinion ${O}_{i}(t) \in [0,1]$ (real numbers between $0$ and $1$) at time $t$, where $0$ and $1$ are the two extremes of conflicting opinions, e.g., conservatism and progressivism. One should be aware that this number does not correspond to a certain merit or superiority. Initially, $O_i(t=0)$ is assigned as a randomly generated number from the uniform distribution in the interval $[0,1]$. The Wikipedia article also has its own opinion $W(t)$ at time $t$, which is the overall stance of Wikipedia on the topic. We set $W(t=0) = 1$, to get the insights of the situation that agents and the Wikipedia article adjust their opinions to the most radical one. Similar to the fact that it is impossible to gauge the ``stance'' of the article and agents to the topic, we do not explicitly display the values and those are used only for stochastic simulation.

For each time step $t$, the agent-agent interaction described in Sec.~\ref{sec:agent_agent_dynamics} and the agent-Wikipedia dynamics described in Sec.~\ref{sec:agent_wiki_dynamics} occur in turn.

\begin{figure*}
\includegraphics[width=1.0\textwidth]{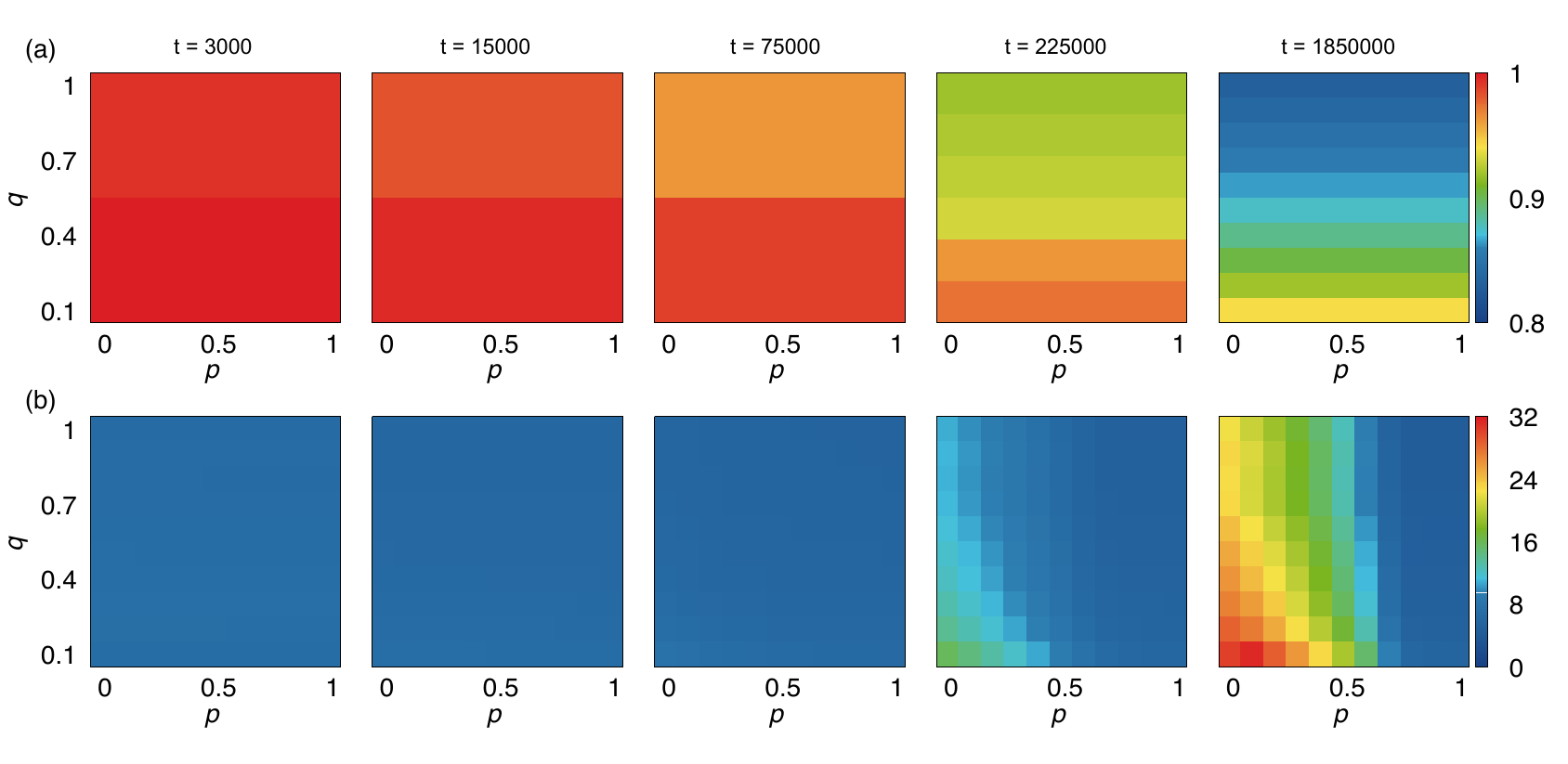}
\caption{Snapshots of (a) $\langle {N}_{p}^{*}(i) \rangle / \langle {N}_{e}^{*}(i) \rangle$ and (b) $\langle S^*(i)\rangle /\langle {N}_{p}^{*}(i)\rangle $, where $\langle \cdots \rangle$ is the averaged quantity over all of the agents ($ = \sum_i \cdots / N$) at five time points from $t = 3000$ to $1850000$. $N = 10^5$ agents are connected by the Erd\H{o}s-R{\'e}nyi random network~\cite{Erdos1960} with connecting probability $0.1$. Although we present the results from the random network case, our results hold for other networks, e.g., the static scale-free networks~\cite{Goh2001}, as well as the fully connected network case. Each point is the average of $1000$ independent runs of simulation with $\epsilon_w = \epsilon_m = \epsilon_l = 0.1$. (a) $\langle {N}_{p}^{*}(i) \rangle / \langle {N}_{e}^{*}(i) \rangle$, which corresponds to the proportionality constant in Fig.~\ref{Bimodality}(a). We observe the ratio gradually falls into different regimes that are determined by $q$ as time passes. The plots in panel (b) represent $\langle S^*(i)\rangle /\langle {N}_{p}^{*}(i)\rangle $ corresponding to Fig.~\ref{Bimodality}(b).}
\label{Modelresults}
\end{figure*}

\subsubsection{Agent-agent dynamics}
\label{sec:agent_agent_dynamics}
Our model colligates resolving of conflicts between agents with the contribution of agents to modify Wikipedia. In our model, all members of society are open-minded and they can change their mind. For each timestep $t$, a pair of agents $i$ and $k$, which are neighbors in a preassigned network, is chosen and they try to convince each other for the topic of the Wikipedia article. We assume that agents rely on references to reinforce their opinion. For simplicity, we consider only two major types of references: Wikipedia and general media. General media, denoting the entire set of references other than Wikipedia, represent the ordinary viewpoint of the society toward the topic. As we described above, Wikipedia is a more reliable source for certain topics. Hence, we set a probability $p$ with which agents choose Wikipedia as their reference, and this probability corresponds to the reliability of Wikipedia [see Fig.~\ref{Modeldescription_fig}(a)]. Otherwise, agents decide to follow the standards of society by following general media's opinion, which is defined as the average opinion of entire agents in the society. In other words, the reference opinion 
\begin{gather}
R(t) = \left\{ \begin{array}{ll}
W(t) & \textrm{with probability $p$}\\
\displaystyle G(t) = \frac{1}{N} \sum_s O_s(t) & \textrm{with probability $1-p$} \,.
\end{array} \right.
\end{gather}

Once we choose the reference, an agent whose opinion is closer to the reference always succeeds in convincing the other agent. For the convenience, we call the agent as $i$ whoever's opinion is closer to the reference than the other, i.e., $|O_i(t) - R(t)| \le |O_k(t) - R(t)|$ [see Fig.~\ref{Modeldescription_fig}(a)]. Agent $k$ changes its opinion toward $i$'s, while agent $i$ keeps its opinion. People tend to minimize the amount of changing~\cite{Sarnoff1960}; thus the agent $k$ sets his/her target as $R(t)$ or $O_i(t)$, depending on which one is closer. As a result, the opinions of agents $i$ and $k$ at the next step $t+1$ are given by
$O_i(t+1) = O_i(t)$ and 

\begin{gather}
{O}_{k}(t+1) = 
\left\{ \begin{array}{ll} 
(1 - \epsilon_m) {O}_{k}(t) + \epsilon_m O_i (t) & \\
\textrm{if $|O_i(t) - O_k(t)| \le |O_k(t) - R(t)|$} & \\
 & \\
(1 - \epsilon_m) {O}_{k}(t) + \epsilon_m R(t) & \\
\textrm{otherwise,} & \\
\end{array} \right.
\end{gather}
respectively. The parameter $\epsilon_m \in [0,1]$ represents the tolerance of agents, which indicates the psychological limit to change the opinion, as discussed in the introductory part of Sec.~\ref{sec:model}. The value of $\epsilon_m$ affects mainly the timescale of simulation, yet does not have a large impact on our model conclusions; thus we fix this value as $0.1$ to set a moderate time scale. 

\subsubsection{Agent-Wikipedia dynamics}
\label{sec:agent_wiki_dynamics}
A distinct character of our model, compared to opinion spreading models with external field~\cite{Castellano2009}, is that agents can also modify the media, which corresponds to Wikipedia in our model. One additional key difference from the previous models of Wikipedia~\cite{Gandica2015, Iniguez2014} is that we introduce the \emph{length} $L(t)$ into our model. This length not only has an impact on the edits but also changes by the edits. We focus on the length of the article instead of the specific opinion values. 

As we described in the introductory part of Sec.~\ref{sec:model}, we assume that there are fundamental differences in ``editablity'' among articles, thus we set the base activity of edit $q$ as the control parameter. For every timestep, a randomly chosen agent attempts to edit the article in Wikipedia. Suppose that agent $e$ is chosen, then it edits the Wikipedia article with the probability

\begin{gather}\label{prob_edit_wiki}
{r}_{e}(t) = {q} |W(t) - O_e(t)| {\frac{A_e(t) + 1}{\sum_{s}{A_s(t) + 1}}}  {\frac {1}{ L(t)}} \,,
\end{gather}

where ${q}$ is base activity, and $A_e(t)$ is attachment of agent $e$ for the Wikipedia article~\cite{George2004}, which is assigned as $0$ at the beginning and increased by unity every time an agent edits the Wikipedia article [see Fig.~\ref{Modeldescription_fig}(b)]. Naturally, the term $[A_e(t) + 1] / [\sum_{s}{A_s(t) + 1}]$ accounts for the fact that people tend to edit more frequently when they have contributed to establishing the current state of the article more~\cite{George2004}. The term ${{1}/{ L(t)}}$ in Eq.~\eqref{prob_edit_wiki} represents the reduced motivation as the article size is increased, due to the amount of information~\cite{Bryant2005}. A recent report that the growth of Wikipedia has slowed down supports this factor~\cite{Suh2009}. If an agent decides to edit an article, Wikipedia's opinion changes as 
\begin{gather}
W(t+1) = W(t) + \epsilon_w [O_e(t) - W(t)] / L(t) \,.
\end{gather}
In our model, the amount of change is inversely proportional to $L(t)$. Figure~\ref{DeltasizeperSize} indicates that the impact of a single edit event should be decreased as the article size is increased, because the absolute amount of change is preserved.  Additionally, $\epsilon_w$ represents the physical and psychological limit for editing~\cite{Gherardi2013}. The value of $\epsilon_w$ affects also mainly the timescale of simulation similar to $\epsilon_m$. We fix this value as $0.1$ to set a moderate time scale, and this value does not have a large impact on our model conclusions.

Finally, the length parameter $L(t)$ is changed after the update of the article's opinion, as follows:
\begin{gather}
L(t+1) = L(t) + C_l(t) \,,
\end{gather}
where the random variable $C_l(t)$ is chosen according to the following rule. If the agent has modified the article toward an extreme position ($0$ or $1$), we suppose that the agent tend to append new contents to the article. In contrast, agents are likely to replace the contents to neutralize the article's opinion. Specifically, we divide the update into the two following cases: (i) $O_e(t) \le W(t) \le 1/2$ or $O_e(t) \ge W(t) \ge 1/2$ (toward an extreme) and (ii) any other cases (neutralize). For (i), the article size is increased by $C_l(t)$ drawn from the interval $[0, \epsilon_l]$ uniformly at random, to reinforce the argument. Otherwise, $C_l(t)$ is drawn from the interval $[-\epsilon_l/2, \epsilon_l/2]$ uniformly at random, which implies replacement of arguments. The fixed parameter $\epsilon_l$ is related to the physical limit in Fig.~\ref{DeltasizeperSize}. The value of $\epsilon_l$ affects mainly the length. However, in this study, we use the ratio of the length to other measures rather than the absolute length of the article. We display the result with $\epsilon_l = 0.1$ (Fig.~\ref{Modelresults}), yet we verify that our conclusions are robust for other values of $\epsilon_l$ because the parameter governs only the overall length scale [see Fig.~\ref{Modeldescription_fig}(b) for the illustration on the $C_l(t)$ criterion]. In Sec.~\ref{sec:model_results}, we discuss how the modes in Fig.~\ref{Bimodality} are formulated during the evolution of Wikipedia in our model.

\begin{figure}
\includegraphics{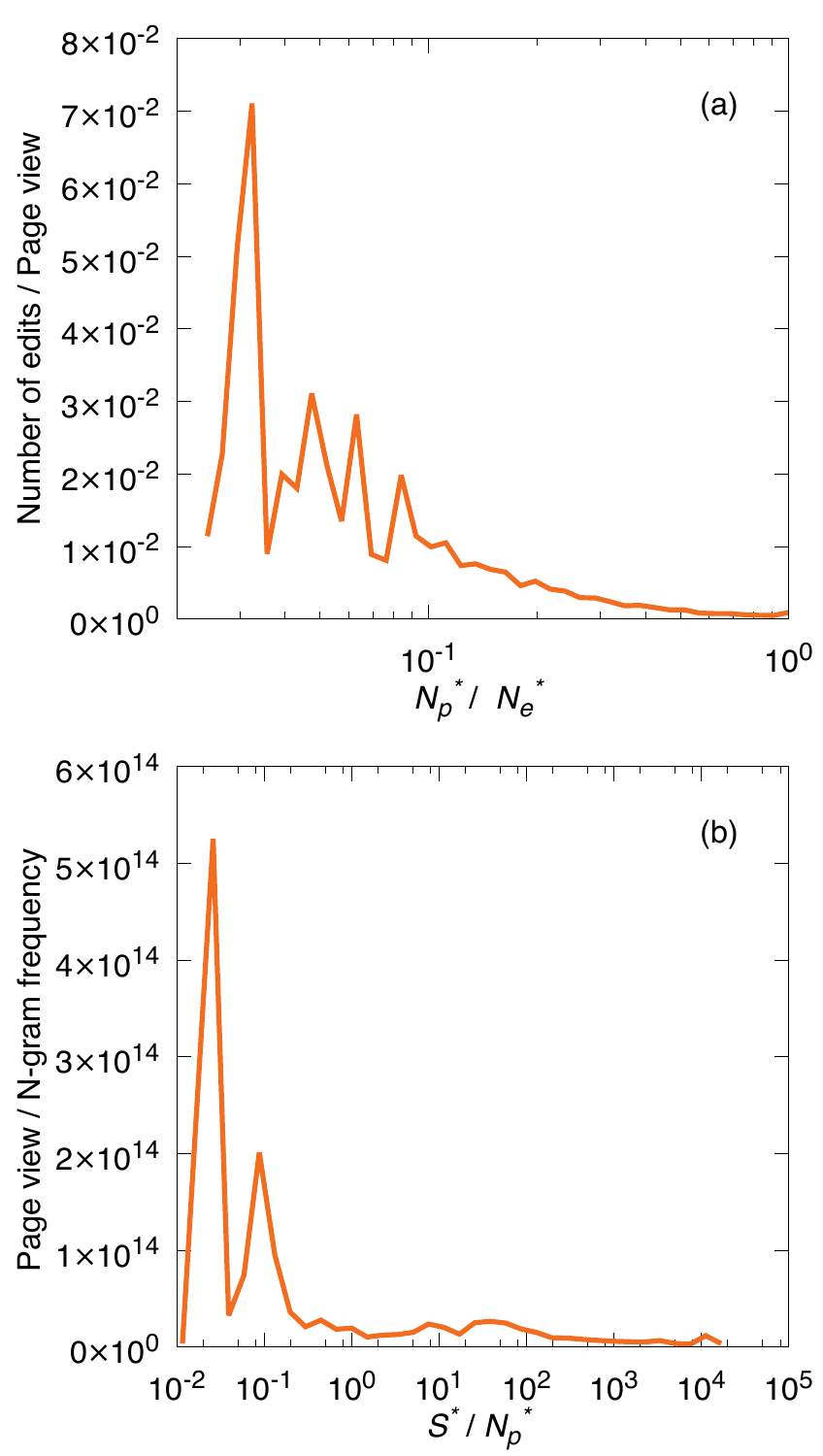}
\caption{Estimation of $q$ and $p$ using the page view statistics of Wikipedia (2014) and Google $n$-gram data set (2008), which are the latest data sets for both.  (a) The average value of estimated $q$, which is calculated by dividing the number of edits in 2014 by the page view statistics in 2014~\cite{WikiTraffic}, as a function of ${N}_{p}^{*}/{N}_{e}^{*}$. As expected, the estimated $q$ decreases for larger ${N}_{p}^{*}/{N}_{e}^{*}$ values. (b) The average value of estimated $p$: the ratio of the page view statistics in 2014~\cite{WikiTraffic} to the Google 1-gram frequency in 2008~\cite{Michel2011}, as a function of $S^{*}/N_p^{*}$. Both plots are drawn from the same sampled set of $678\,255$ articles, with the conditions described in the text. 
}
\label{Modelverify}
\end{figure}

\begin{figure*}
\includegraphics{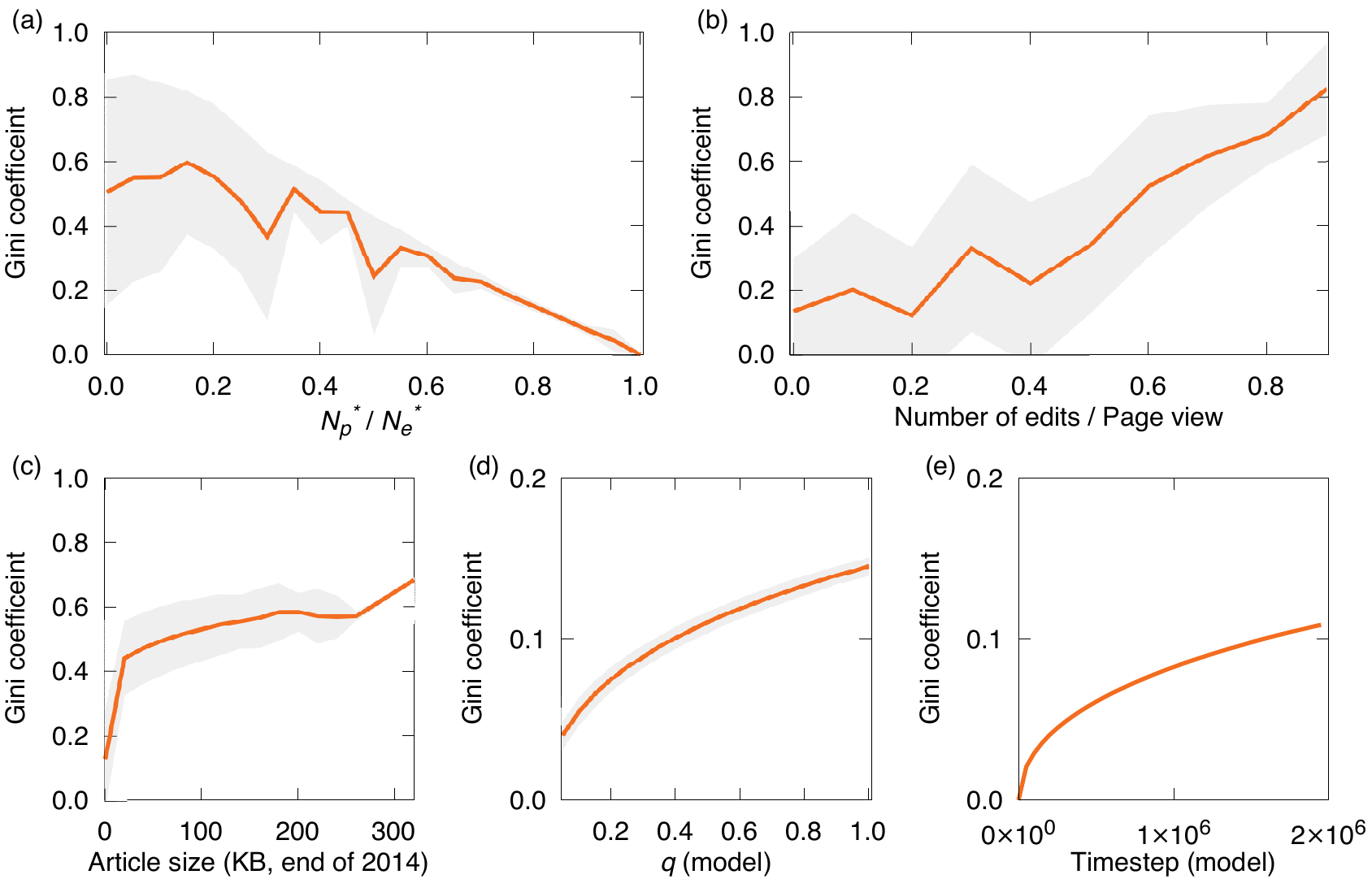}
\caption{The Gini coefficient values for Wikipedia articles classified by various measures. Each sample in the data points corresponds to an individual Wikipedia article.  The average coefficient is plotted as functions of (a) ${{N}_{p}^{*}}/{{N}_{e}^{*}}$, (b) the estimator of $q$: ratio of the number of edits in 2014 to the page view statistics, and (c) the article size at the end of 2014. Panels (d) and (e) display the observation from our model, as functions of (d) $q$ and (e) the timestep. For panels (a)--(d), the shaded areas correspond to the standard deviation of Gini coefficient for given values on the horizontal axis. Panels (a)--(c) with the Wikipedia data are drawn from the same sampled set of $678\,255$ articles used in Fig.~\ref{Modelverify}. The model results in (d) and (e) are also obtained using the same condition as in Fig.~\ref{Modelresults}, but with the fixed $p = 0.5$.
}
\label{Gini}
\end{figure*}

\subsection{Model results}
\label{sec:model_results}
\subsubsection{Conditions for bimodality}
\label{sec:conditions_for_bimodality}
For both ${N}_{e}^{*}(i)$ versus ${N}_{p}^{*}(i)$ and ${N}_{p}^{*}(i)$ versus ${S}^{*}(i)$ relations, our mechanistic model captures the essential features of the observed empirical relations reported in Sec.~\ref{sec:time_rescaled} with proper parameter values. As we have shown in Sec.~\ref{sec:data_analysis}, the proportionality coefficients between characteristic parameters are classified into two modes: 
\begin{gather}
{{N}_{p}^{*}(i)}/{{N}_{e}^{*}(i)} ~ \simeq~0.5~\textrm{or}~0.9\\
{{S}^{*}(i)}/{{N}_{p}^{*}(i)} ~ \simeq~25~\textrm{or}~900
\end{gather}
In particular, both $p$ for the agent-agent interaction (in Sec.~\ref{sec:agent_agent_dynamics}) and $q$ for the agent-Wikipedia interaction (in Sec.~\ref{sec:agent_wiki_dynamics}) are crucial to generate the splits of modes into different groups: $q$ is essential to reproduce a separation of ${{N}_{p}^{*}(i)}/{{N}_{e}^{*}(i)}$ [see Fig.~\ref{Modelresults}(a)] and $p$ is indispensable for the division of ${{S}^{*}(i)}/{{N}_{p}^{*}(i)}$ [see Fig.~\ref{Modelresults}(b)]. In the early stage, ${{N}_{p}^{*}(i)}/{{N}_{e}^{*}(i)}$ is almost unity across the systems with the entire parameter space composed of $p$ and $q$, which corresponds to the single (or unimodal, in contrast to the bimodal pattern shown in real data) linear relation. While this single linear relation is characterized at the early stage, as time goes by, we observe the decreasing trend of ${{N}_{p}^{*}(i)}/{{N}_{e}^{*}(i)}$. 

Despite the fact that the ${{N}_{p}^{*}(i)}/{{N}_{e}^{*}(i)}$ decrement over time occurs for the entire parameter space, the pace of decreasing is determined by $q$, the base rate for editing an article. ${{N}_{p}^{*}(i)}/{{N}_{e}^{*}(i)}$ drops much slower for smaller $q$ values, which leads systems to fall into two different regimes: $q \gtrsim 0.5$ and $q \lesssim 0.5$ [Fig.~\ref{Modelresults}(a)]. Interestingly, this divarication solely depends on the value of $q$. On the other hand, ${{S}^{*}(i)}/{{N}_{p}^{*}(i)}$ also shows unimodality in the early stage, but it is suddenly increased with time only for $p \lesssim 0.5$ [Fig.~\ref{Modelresults}(b)]. Analogous to ${{N}_{p}^{*}(i)}/{{N}_{e}^{*}(i)}$, ${{S}^{*}(i)}/{{N}_{p}^{*}(i)}$ is also almost solely driven by $p$, but there also exists a small amount of influence by $q$; small values of $q$ do not guarantee the large article size across all $p$ values, but low $q$ yields large article size at similar values of $p$.

\subsubsection{Model verification with real data}

Based on our model results, we suggest a possible mechanism that yields the bimodality in Fig.~\ref{Bimodality}, which encourages us to verify the model results compared to the real data: either parameter $q$ or $p$ should be a decreasing function of ${{N}_{p}^{*}}/{{N}_{e}^{*}}$ and ${{S}^{*}}/{{N}_{p}^{*}}$, respectively. However, we cannot extract simulation parameters $p$ and $q$ directly from the data. We therefore use a bypass to estimate $q$ and $p$. Fortunately, Wikipedia offers page view statistics of articles that can be used for estimating such parameters~\cite{WikiTraffic}. We assume that this page view in a certain period reflects the degree of interest of Wikipedia users in the articles, and the number of edits in the same period naturally displays the editing frequency. Thus, the ratio of the number of edits to this page view for a certain period can be related to the base edit rate $q$. Analogous to our presumption, this ratio is a decreasing function of ${{N}_{p}^{*}}/{{N}_{e}^{*}}$ (see Fig.~\ref{Modelverify}). 

To treat the other parameter $p$, we should employ the proxy that can reflect the general interest of the entire society in the topic. We suggest that the Google books $n$-gram, a vast digitized collection of documents produced in the world is a suitable choice~\cite{Michel2011, Yun2015}. Google books $n$-gram is a database containing about 6\% of English books ever published. This data set offers a yearly number of occurrences for any phrase less than six words from 1800 to 2008, and this number of occurrences can be considered as the proxy of interests in society for a certain phrase. In our model, $p$ is the proportion of degree of interest in Wikipedia versus that of the entire society. In other words, Wikipedia page view on a certain topic versus its $n$-gram frequency can be the estimator of $p$. For fair comparison, we also only take the Wikipedia articles that satisfy the following conditions. First, the title of article should exist in Google 1-gram data set in 2008, the latest year of the data set. Second, the article should be visited at least once in 2014. To avoid the effect of inflectional variation of words, we use the stem of Wikipedia articles' title and Google 1-gram data set~\cite{Porter1980}, instead of using the word directly. After this filtering process, $678\,255$ articles are left among the total set of $34\,534\,110$ articles. This estimator of $p$ also decays, as ${{S}^{*}}/{{N}_{p}^{*}}$ is increased as we expected. 

Both Figs.~\ref{Modelverify}(a) and \ref{Modelverify}(b) indeed show the behaviors expected from their estimators ($q$ and $p$, respectively), which indicates that our model is suitable to describe the real Wikipedia. Note that the estimators of $q$ and $p$ should not be taken as the exact face values of model parameter values for a real article in Wikipedia, and the results should be understood as a proxy of statistical properties of articles. First, page view statistics might be affected by the number of hyperlinks pointed to the article. Such relative importance within the network topology may increase the page view by random visits, yet there is a positive feedback between the page view and the number of hyperlinks. An article also tends to have connections to popular articles~\cite{Capocci2006}, which eventually yields disproportionally many hyperlinks for popular items; thus there could be overestimation of page views for the popular articles. Moreover, there is a recent report that warns of the possible bias of Google $n$-gram as the proxy of real popularity in our society~\cite{Pechenick2015}. This year-by-year level fluctuation may give unfairness to compare the frequencies many years apart. To avoid such fluctuation, we restrict our results for the year 2008. Additionally, word-by-word fluctuation should be canceled during the averaging process, because each data point corresponds to a massive number of articles. As a result, we believe that our observation is still valid, in spite of such fluctuations that might cause some degree of bias.

\subsubsection{Inequalities in Wikipedia editing}
The heterogeneity for the ratio of the number of editors to that of edits, ${{N}_{p}^{*}}/{{N}_{e}^{*}}$, leads us to the eventual question: is this heterogeneity from structural inequality? In other words, does the existence of dictatorship or monopoly of small group editors, or super editors~\cite{Gandica2015}, make it difficult for others to participate in editing processes? To find the answer, we use the Gini coefficient, which is a common measure for inequality in economics~\cite{Gini1912} ranging from $0$ for the minimal inequality (or the maximal equality) to $1$ as the maximal inequality. We consider the number of edits for individual editors as the wealth variable in the Gini coefficient. The trend of the Gini coefficient as a decreasing function of ${{N}_{p}^{*}}/{{N}_{e}^{*}}$ shown in Fig.~\ref{Gini}(a) suggests the modes with slope $\simeq 0.9$ and $\simeq 0.5$ in Fig.~\ref{Bimodality}(a) are in equilibrium and non-equilibrium states, respectively. 

Additional analysis of the Gini coefficient in terms of the $q$ estimator (the ratio of the number of edits in 2014 to the page view statistics in 2014) also indicates that the larger $q$ induces more severe inequality for editing [see Fig.~\ref{Gini}(b)]. This is counterintuitive because it actually means that articles inducing larger motivation to edit eventually set a larger barrier to participate in editing. It is doubtful that the phenomenon is caused by the amount of information~\cite{Bryant2005}, since the Gini coefficient does not vary much according to its amount of information [see Fig.~\ref{Gini}(c)]. 

Similar to the real Wikipedia, our model also supports the observed inequality. Although we use a simplified estimator of $q$ in our real data, the ratio of the number of edits in 2014 to the page view statistics, the Gini coefficient is an increasing function of $q$ in the model as in the real data [see Fig.~\ref{Gini}(d)]. Additionally, since $q$ has a limited effect on the article size (see Fig.~\ref{Bimodality}), the model observation of the Gini coefficient is compatible to our observation that article size does not have a large effect on the Gini coefficients. Such logical elimination suggests that a few engaged and dominating editors make it indeed hard for laypeople to participate in editing processes. There are ``democratic'' articles (with slope of $\simeq 0.9$ in Fig.~\ref{Bimodality}) and ``dictatorial'' articles (with slope of $\simeq 0.5$ in Fig.~\ref{Bimodality}). In short, inequality exists indeed. 
 
\section{Conclusion}
\label{sec:conclusion}
Traditionally, collaboration used to be mainly regional and face-to-face interactions were demanded, which had prevented the world-wide formation of collective intelligence. Nowadays, improvements of modern information technology bring us a whole new stage of online collaboration. In this study, we have examined such a new passion of collective intelligence through long-term data from Wikipedia~\cite{Wikipedia}. People believe that such a new paradigm will eventually yield democratization of knowledge~\cite{Lemke2009}. As a representative medium, Wikipedia is also considered as a spearhead of such pro-democracy movements~\cite{Hasan2006}. 

However, our observation suggests that the current status of Wikipedia is still apart from the perfect world-wide democracy. The observed periodicity for the time between edits alludes that the English Wikipedia is still regional for English natives (see Fig.~\ref{DeltatimeperAge}). Bimodality and its inequality index suggest that there are articles dominated by a small number of super editors (Figs.~\ref{Bimodality}--\ref{Gini}). Notwithstanding the fact that there is no explicit ownership for Wikipedia articles, some kind of privatization by dedicated editors for given topics is happening in reality. The value of such dedicated editors should not be depreciated, of course. Their dedication has indeed played the main role in keeping the current state-of-the-art accuracy in the current Wikipedia~\cite{Chesney2006, Giles2005}. However, in the long run, knowledge cannot survive without collaboration between experts and society~\cite{Yun2015}. Although most advanced knowledge is invented by experts, such experts occupy a rather small proportion in a society; thus, knowledge without support from other members of the society will lose its dynamic force to sustain. Additionally, despite our findings that the amount of contents created by an editor (${{S}^{*}}/{{N}_{p}^{*}}$) mainly depends on the degree of referring Wikipedia (namely $p$), an equitable opportunity for participation also increases such individual productivity (see Fig.~\ref{Modelresults}).

Our study not only gives significant insight into the formation and current state of Wikipedia, but also offers the future direction of Wikipedia. Our simulation results suggest that such inequality is increased with time, which may result in less productivity and less accuracy as by-product in the future than now [see Figs.~\ref{Modelresults}(a) and \ref{Gini}(e)]. It is indeed already reported that the growth of Wikipedia is slowing down~\cite{Suh2009} and our observation suggests that it will become even slower if we do not take any active action. To sustain collaborating environments, it is worth giving more motivation and incentives to the newbies to reduce the monopolized structure in Wikipedia. We hope that extending our approach to various collaboration environments such as open-source movement~\cite{Lakhani2005, Lerner2002} might give us the insight for the future investment that brings us a new level of collaborating environments. Finally, we would like to emphasize that the results and implications of our study are not restricted to the Wikipedia or online collaboration systems, but have much wider applications in human or nonhuman interactions in the world.

\begin{acknowledgments}
We are grateful to Beom Jun Kim (김범준), Pan-Jun Kim (김판준), and Hyunggyu Park (박형규) for insightful comments. This work was supported by the National Research Foundation of Korea through Grant No. 2011-0028908 (J.Y. and H.J.). 
\end{acknowledgments}


\end{CJK}

\begin{thebibliography}{00}

\bibitem{Pliny77} Pliny the Elder, \emph{Naturalis Historia (Natural history)} (Ancient Rome, AD77--79).

\bibitem{Lindberg2007} D. Lindberg. \emph{The Beginnings of Western Science} (University of Chicago Press, Chicago, 2007).

\bibitem{Britannica} \emph{Encyclopaedia Britannica, 15th ed.} (Encyclop{\ae}dia Britannica, Inc., Chicago, 2010).

\bibitem{Americana}  \emph{Encyclopedia Americana, Intl. ed.} (Scholastic Library Publishing Inc.,  Danbury, 2005).

\bibitem{BritannicaWeb} \emph{Encyclopaedia Britannica}, Online edition, \url{http://global.britannica.com/}.

\bibitem{Wikipedia} \emph{Wikipedia}, \url{https://www.wikipedia.org/}.

\bibitem{WikipediaDisclamer} \emph{Wikipedia disclamer}, \url{https://en.wikipedia.org/wiki/Wikipedia:General_disclaimer}.

\bibitem{Kittur2008} A. Kittur, B. Suh, and E.\,H. Chi, \emph{Can you ever trust a Wiki?: Impacting perceived trustworthiness in Wikipedia}, Proceedings of the 2008 ACM Conference on Computer Supported Cooperative Work (CSCW '08), p. 477 (2008). 

\bibitem{Adler2008} B.\,T. Adler, K. Chatterjee, L. de Alfaro, M. Faella, I. Pye, and V. Raman,  \emph{Assigning trust to Wikipedia content}, Proceedings of the 4th International Symposium on Wikis (WikiSym '08), Article No. 26 (2008). 

\bibitem{Nielsen2007} F.\,A. Nielsen, \emph{Scientific citations in Wikipedia}, First Monday {\bf 12}, 8 (2007).

\bibitem{Haigh2011} C.\,A. Haigh, \emph{Wikipedia as an evidence source for nursing and healthcare students}, Nurse Educ. Today {\bf 31}, 135 (2011).

\bibitem{Chesney2006} T. Chesney, \emph{An empirical examination of Wikipedia's credibility}, First Monday {\bf 11}, 11 (2006).

\bibitem{Giles2005} J. Giles, \emph{Internet encyclopedias go head to head}, Nature {\bf438}, 900 (2005).

\bibitem{Bould2014} M.\,D. Bould,  E. S. Hladkowicz, A.-A.\,E. Pigford, L.-A. Ufholz, T. Postonogova,  E. Shin, and S. Boet, \emph{References that anyone can edit: review of Wikipedia citations in peer reviewed health science literature}, BMJ {\bf348}, g1585 (2014).

\bibitem{Amichai2008} Y. Amichai-Hamburger, N. Lamdan, R. Madiel, and T. Hayat, \emph{Personality characteristics of Wikipedia members}, Cyberpsychol. Behav. {\bf11}, 679 (2008).

\bibitem{Iba2010} T. Iba, K. Nemoto, B. Peters, and P.\,A. Gloord, \emph{Analyzing the creative editing behavior of Wikipedia editors: Through dynamic social network analysis}, Procedia Soc. Behav. Sci. {\bf2}, 6441 (2010).

\bibitem{Gandica2015} Y. Gandica, J. Carvalho, and F. Sampaio dos Aidos, \emph{Wikipedia editing dynamics}, Phys. Rev. E {\bf91}, 012824 (2015).

\bibitem{Yasseri2012} T. Yasseri, R. Sumi, A. Rung, A. Kornai, and J. Kert{\'e}sz, \emph{Dynamics of conflicts in Wikipedia}, PLOS ONE {\bf7}, e38869 (2012).

\bibitem{Iniguez2014} G. Iniguez, J. Torok, T. Yasseri, K. Kaski, and J. Kert{\'e}sz, \emph{Modeling social dynamics in a collaborative environment}, EPJ Data Sci. {\bf3}, 7 (2014).

\bibitem{WikimediaDownloads} \emph{Wikimedia downloads}, \url{https://dumps.wikimedia.org/backup-index.html}.

\bibitem{Utf8} F. Yergeau \emph{UTF-8, a transformation format of ISO 10646} STD 63, RFC 3629.

\bibitem{BroadbandShare} \emph{Broadband Internet penetration deepens in US; Cable is king}, \url{https://technology.ihs.com/468148/broadband-internet-penetration-deepens-in-us-cable-is-king} (2013).

\bibitem{YXie2007} Y. Xie, F. Yu, and K. Achan, \emph{How dynamic are IP addresses?}, Proc. ACM SIGCOMM'07 (2007)

\bibitem{Goh2008} K.-I. Goh and A.-L. Barab{\'a}si, \emph{Burstiness and memory in complex systems}, EPL {\bf81}, 48002 (2008).

\bibitem{Barabasi2005} A.-L. Barab{\'a}si, \emph{The origin of bursts and heavy tails in human dynamics}, Nature {\bf435}, 207 (2005).

\bibitem{Kittur2007} A. Kittur, B. Suh, B. A. Pendleton, and E.\,H. Chi, \emph{He says, she says: Conflict and coordination in Wikipedia}, Proceedings of the SIGCHI Conference on Human Factors in Computing Systems (CHI'07), p. 453 (2007).

\bibitem{Yasseri2012a} T. Yasseri, R. Sumi, and J. Kert{\'e}sz, \emph{Circadian patterns of Wikipedia editorial activity: A demographic analysis}, PLOS ONE {\bf 7}, e30091 (2012).

\bibitem{Gherardi2013} M. Gherardi, S. Mandr{\`a}, B. Bassetti, and M.\,C. Lagomarsino, \emph{Evidence for soft bounds in Ubuntu package sizes and mammalian body masses}, Proc. Natl. Acad. Sci. USA {\bf110}, 21054 (2013).

\bibitem{Taneska2009} B.\,K. Taneska, \emph{Otaku---The living force of the social media network}, Third International Communication Conference -- Glocal: Inside Social Media, 2009 (unpublished). 

\bibitem{Bryant2005} S.\,L. Bryant, A. Forte, and A. Bruckman, \emph{Becoming Wikipedian: Transformation of participation in a collaborative online encyclopedia}, Proceedings of the 2005 international ACM SIGGROUP Conference on Supporting Group Work (GROUP '05) (2005).

\bibitem{Finkel1998} S.\,E. Finkel and E.\,N. Muller, \emph{Rational choice and the dynamics of collective political action: Evaluating alternative models with panel data}, Am. Polit. Sci. Rev. {\bf92}, 37 (1998).

\bibitem{Lakhani2005} K.\,R. Lakhani and R. Wolf, Why hackers do what they do: Understanding motivation and effort in free/open source software projects, \emph{Perspectives on Free and Open Source Software} (MIT Press, Cambridge, MA, 2005).

\bibitem{George2004} B.\,P. George, \emph{Past visits and the intention to revisit a destination: Place attachment as the mediator and novelty seeking as the moderator}, J. Tour. Stud. {\bf15}, 51 (2004).

\bibitem{Erdos1960} P. Erd\H{o}s  and A . R{\'e}nyi, \emph{On the evolution of random graphs}, Publ. Math. Inst. Hungar. Acad. Sci. {\bf5}, 17 (1960).

\bibitem{Gilbert1959} E.\,N. Gilbert, \emph{Random graphs}, Ann. Math. Stat. {\bf30}, 1141 (1959).

\bibitem{Sarnoff1960} I. Sarnoff, \emph{Psychoanalytic theory and social attitudes}, Public Opin. Quart. {\bf24}, 251 (1960).

\bibitem{Castellano2009} C. Castellano, S. Fortunato, and V. Loreto, \emph{Statistical physics of social dynamics}, Rev. Mod. Phys. {\bf81}, 591 (2009).

\bibitem{Suh2009} B. Suh, \emph{The singularity is not near: Slowing growth of Wikipedia}, Proceedings of the 5th International Symposium on Wikis (WikiSym '09), Article No. 8 (2009). 

\bibitem{Goh2001} K.-I. Goh, B. Kahng, and D. Kim, \emph{Universal behavior of load distribution in scale-free networks}, Phys, Rev. Lett. {\bf87}, 278701 (2001).

\bibitem{WikiTraffic} \emph{Page view statistics for Wikimedia projects}, \url{http://dumps.wikimedia.org/other/pagecounts-raw/}.

\bibitem{Michel2011} J.-B. Michel {\it et al.}, \emph{Quantitative analysis of culture using millions of digitized books}, Science {\bf331}, 176 (2011). 

\bibitem{Yun2015} J. Yun, P.-J. Kim, and H. Jeong, \emph{Anatomy of scientific evolution}, PLOS ONE {\bf10}, e0117388 (2015).

\bibitem{Porter1980} M.\,F. Porter, \emph{An algorithm for suffix stripping}, Program {\bf14}, 130 (1980).

\bibitem{Capocci2006} A. Capocci, V.\,D.\,P. Servedio, F. Colaiori, L.\,S. Buriol, D. Donato, S. Leonardi, and G. Caldarelli, \emph{Preferential attachment in the growth of social networks: The internet encyclopedia Wikipedia}, Phys. Rev. E {\bf74}, 036116 (2006).

\bibitem{Pechenick2015} E.\,A. Pechenick, C.\,M. Danforth, and P.\,S Dodds, \emph{Characterizing the Google Books corpus: Strong limits to inferences of socio-cultural and linguistic evolution}, PLOS ONE {\bf 10}, e0137041 (2015).

\bibitem{Gini1912} C. Gini, \emph{Variabilit{\`a} e mutabilit{\`a} (Variability and mutability)} (C. Cuppini, Bologna, 1912).

\bibitem{Lemke2009} C. Lemke and E. Coughlin, \emph{The change agents}, Teaching for the 21st Century {\bf67}, 54 (2009).

\bibitem{Hasan2006} H. Hasan and C. Pfaff, \emph{Emergent conversational technologies that are democratising information systems in organisations: The case of the corporate Wiki}, Proceedings of the Information Systems Foundations (ISF): Theory, Representation and Reality Conference (2006).

\bibitem{Lerner2002} J. Lerner and J. Triole, \emph{Some simple economics of open source}, J. Ind. Econ. {\bf50}, 97 (2002).

\end{thebibliography}
\end{document}